%% file: paper.tex
\documentclass{sig-alternate-10pt}

\usepackage[font=small,format=plain,labelfont=bf,up,textfont=bf,up]{caption}
\usepackage{times}
\usepackage{verbatim}
\usepackage{ifthen}
\newboolean{TECHREPORT}
\setboolean{TECHREPORT}{true}
\usepackage{amsmath}
\usepackage{algorithm}
\usepackage{algorithmic}
\usepackage{multirow}

\usepackage{graphicx}
\usepackage{endnotes,url}
\usepackage[sort]{cite}
\usepackage{subfigure}
\usepackage{calc}
\usepackage{verbatim,amsfonts,amssymb}
\usepackage{multirow}
\usepackage{epsfig}
\usepackage{xspace}
\usepackage{mdwlist}
\usepackage[dvips,colorlinks=true,citecolor=black,linkcolor=black,urlcolor=black]{hyperref}
\usepackage[hyphenbreaks]{breakurl}
\usepackage[usenames]{color}

\usepackage{soul}

\usepackage [titletoc] {appendix}

\newcommand{\ignore}[1]{}
\newcommand{\spacesave}[1]{}

\newcommand{\sys}{SybilFence\xspace}

\newcommand{\fp}{\vspace*{0.05in}\noindent}
\renewcommand{\paragraph}[1]{\fp {\bf #1}~}

\begin{document}

\title{SybilFence: Improving Social-Graph-Based Sybil Defenses with User Negative Feedback}


\author{
\begin{tabular}{c}
Qiang Cao ~~~~~~~~~~~~~Xiaowei Yang \\
Duke University\\
\{qiangcao, xwy\}@cs.duke.edu \\
\end{tabular}
}

\maketitle

\input{abstract}
\input{intro}

\input{motivating-observation}
\input{system-design}

\input{evaluation}

\input{related-work}

\input{conclusion}

\let\oldthebibliography=\thebibliography
\let\endoldthebibliography=\endthebibliography
\renewenvironment{thebibliography}[1]{%
    \begin{oldthebibliography}{#1}%
    \setlength{\parskip}{0ex}%
    \setlength{\itemsep}{0ex}%
}{\end{oldthebibliography}}

{\scriptsize \bibliographystyle{abbrv} \bibliography{wosn-12}}

\input{appendix}

\end{document}

%% file: abstract.tex
\begin{abstract}

Detecting and suspending fake accounts (Sybils) in online social networking (OSN) services 
protects both OSN operators and OSN users from illegal exploitation. 
Existing social-graph-based defense schemes effectively bound the accepted Sybils to the total number
of social connections between Sybils and non-Sybil users.
However, Sybils may still evade the defenses by soliciting many social connections to real users. 
We propose \sys, a system that improves over social-graph-based Sybil defenses to further thwart Sybils. 
\sys is based on the observation that even well-maintained fake accounts inevitably receive a 
significant number of user negative feedback, such as the rejections to their friend requests.
Our key idea is to discount the social edges on users that have received negative feedback, 
thereby limiting the impact of Sybils' social edges. 
The preliminary simulation results show that our proposal is more resilient to attacks where
fake accounts continuously solicit social connections over time. 





\end{abstract}

%% file: intro.tex
\section{Introduction}

The popularity of online social networking (OSN) services such as Facebook and LinkedIn has 
attracted attacks and exploitation.
In particular, OSNs are vulnerable to Sybil attacks, where
attackers create many fake accounts, called {\it Sybils}, to send spam~\cite{Gao-IMC-10}, manipulate 
online voting~\cite{Tran-NSDI-09}, crawl users' personal information~\cite{Boshmaf-ACSAC-11}, etc.


There has been several proposals that leverage the underlying social graph to defend against Sybils~\cite{SybilGuard-SIGCOMM-06,
Yu-SP-08, Danezis-NDSS-09, Tran-NSDI-09, Viswanath-SIGCOMM-10, Cao-NSDI-12}. The {\it social-graph-based Sybil defenses}
are proactive approaches, as Sybils can be uncovered before they interact with real users. Those proposals have been extensively discussed in
the research community due to their simplicity and reliability. 
The social-graph-based Sybil defenses rely on an assumption 
that the social edges connecting Sybils and non-Sybil users, called {\it attack edges}, are strictly limited. Most of them
bound the undetectable Sybils, called {\it accepted Sybils}, to the number of attack edges~\cite{Yu-SP-08, Cao-NSDI-12}, 
i.e. $O(\log n)$ Sybils per attack edge.

Although 
social-graph-based Sybil defenses are able to provide theoretical guarantees on accepted Sybils, 
the upper bound of accepted Sybils 
still depends on the total number of attack edges.
Therefore, Sybils have the incentive
to solicit for social connections from real users to increase the attack edges and evade the detection.
Furthermore, {\it well-maintained Sybils} can choose to continuously solicit social edges,
at a speed similar to real users.  
As a result, they may be able to accumulate 
many attack edges from {\it promiscuous real users}, who are open to befriending even strangers.
With current social-graph-based 
Sybil defenses, the fake accounts behind those well-maintained Sybils are indistinguishable from non-Sybil users,
because this entire set of Sybils has adequate connectivity to non-Sybil users.

Fortunately, we observe that the attack edges from 
Sybils 
are usually accompanied by the negative feedback from 
{\it cautious real users}, who are resistant to 
abusive communication. 
A {\it negative feedback} can be a rejection to a friend request or a report on receiving unwanted communication. 
We have been conducting a study on live fake Facebook accounts in the wild (\S\ref{subsec:neg_befriend}),
and find a significant number of negative feedback (pending friend requests) 
on well-maintained fake accounts that are purchased in black market,
although those accounts may manage to connect to real users.

Our understanding of this observation is that the controllers behind the fake accounts have 
limited knowledge about the users' security awareness.
OSN users have varying levels of security 
awareness of the potential exploitation from fake accounts. Promiscuous users have high tolerance
of abusive activities and unwanted communication, while cautious users are more resistant to fake accounts. 
The controller cannot distinguish these types of OSN users, and thus cannot correctly target promiscuous users. 
As a result, fake accounts are likely to receive negative feedback 
from a set of cautious users, although they may be able to interact with some promiscuous users.


To leverage this observation,
we propose \sys, 
which improves over social-graph-based Sybil defenses by incorporating
user negative feedback. Our key idea is to discount the social edges on users who have received
negative feedback. By penalizing the social edges that are accompanied by negative feedback, 
we are able to mitigate the impact of Sybils' attack edges, and construct a {\it defense graph} with reduced weights on attack edges.
In this paper, we use SybilRank, a state-of-the-art social-graph-based Sybil defense scheme~\cite{Cao-NSDI-12}, as a proof of concept, 
and adapt it to the weighted defense graph. 

As shown in 
simulations (\S\ref{sec:eval}), with the defense graph, \sys improves over SybilRank by 
10\%$\sim$20\% in term of the probability of ranking non-Sybil
users higher than Sybils. 
\sys is shown to be more resilient to attacks where well-maintained Sybils keep soliciting for social connections over time.
We conjecture that 
\sys can also improve other Sybil detection schemes such as SybilLimit and Sybil tolerance schemes
such as Bazaar~\cite{post-2011-bazaar}.

%% file: motivating-observation.tex
\section{Sybil Activities and  Negative \\ Feedback}
\label{sec:motivating_observ}

At a high level, the user negative feedback must be triggered by 
abusive activities in OSNs, and reflect the user distrust to the 
executor of the abuse. In practice,
the user negative feedback includes rejections to friend requests, flags on inappropriate
incoming communication such as spam, phishing, pornography and extreme violence, etc.
We observe that such negative feedback
has already existed in OSNs.
OSNs like Facebook have collected and stored such user negative feedback, 
although some of them may be currently 
underutilized or ignored by OSN operators.
Therefore, we do not introduce
any change to current OSNs' use model, and our proposal is completely transparent to users.



We next take the negative feedback triggered by abusively befriending as an example and demonstrate 
why the negative feedback occurs and how the negative feedback associates to Sybil activities.


\subsection{Negative feedback during befriending}
\label{subsec:neg_befriend}
Befriending real users is the first step for fake accounts to infiltrate an OSN
after their creation. During this stage, fake accounts attempt to establish many social connections 
to real users. However, aggressively befriending strangers may trigger negative feedback
from resistant users. For instance,
a bilateral social connection 
requires reciprocal agreements from both users.
The negative feedback during befriending 
can be a rejected or an ignored friend request. 

\paragraph{Study on fake Facebook accounts in black market.}
To better understand the rejection to fake accounts' friend requests in real world,
we have been conducting a study on fake Facebook accounts in black market.
The fake accounts in black market is an example of the well-maintained live fake accounts in the wild.
Those fake accounts
are priced based on the ages, the number of friends, the number of pictures, etc. 
According to our purchase experience, a fake account with 50 $\sim$ 100 friends costs \$2 $\sim$ \$6
at different vendors. Those fake accounts look real and their friends also have rich content in profiles, walls, etc.
We have purchased accounts from different vendors via Freelancer~\cite{Freelancer} and BlackHatWorld~\cite{BlackHatWorld}. 
Apart from pictures, emails, security Q/A sets, we explicitly require in our purchases that the accounts should have
``$>$50 real US friends''. The vendors always ask a relatively long time period to deliver accounts after we order, 
e.g., one week or one month.
Our work is still in progress for a large-scale study on live fake accounts in black market. 

\begin{figure}[t!]
\begin{center}
\epsfig{file=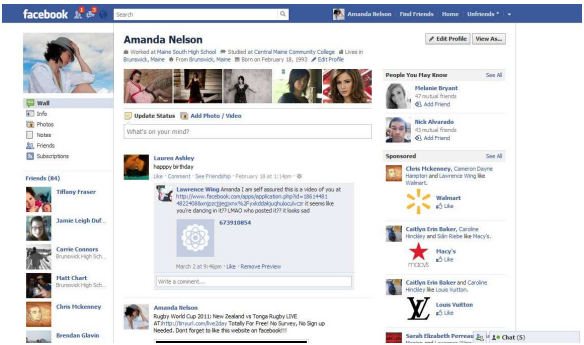, height=4.5cm}
\caption{\bf \footnotesize A sample of purchased accounts. 
\label{fig:fake_sample}}
\end{center}
\end{figure}

\paragraph{Fake accounts look real.} In this study, we use 12 fake accounts with 837 total friends. 
Those accounts are purchased at different vendors.
All of them are at least one year old.
Figure~\ref{fig:fake_sample} is the profile of a sample account. The profile is crafted as a college student with pictures,
and with posts on the wall. This account has 84 friends and has interacted with some friends, such as messages and comments. 


\begin{figure}[t!]
\begin{center}
\epsfig{file=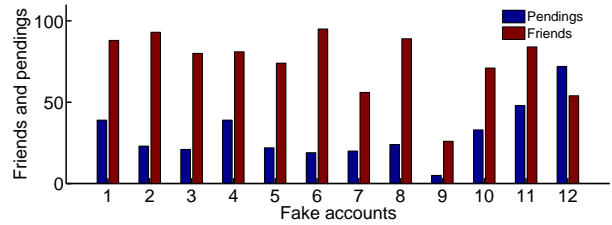, height=2.8cm}
\caption{\bf \footnotesize Friends and pending friend requests on purchased accounts. 
\label{fig:purchase}}
\end{center}
\end{figure}

\paragraph{Fake accounts receive rejections.} 
In Facebook, a pending request implies a rejection, 
as Facebook does not provide the rejection option.
Therefore, we 
examine the pending friend requests on each account. Facebook does not provide this statistic to users directly, but provides APIs to 
access the pending friend requests. Figure~\ref{fig:purchase} shows the number of friends and pending requests on each account.
As we can see, although those well-maintained accounts have many social connections to users that may be real, each of them still has
a significant number of pending requests. 

This result also indicates that the friend requests from those accounts 
have an acceptance rate $>$50\%.
It is reported that randomly flooding friend requests only yields an acceptance rate less
than 30\%~\cite{Boshmaf-ACSAC-11, Panagiotopoulos-SybilFence-11}. We speculate that the vendors might exploit some befriending strategies to
improve the acceptance rate, such as sending requests to the friends of the users that they have already befriended 
(triadic closure principle)~\cite{Boshmaf-ACSAC-11}.


\paragraph{Fake accounts can befriend real users.} To estimate how many real users the fake
accounts have befriended, we choose 2 accounts and send a message to each of their friends. In the message, we 
inform the friends that the account is fake, suggest them to disconnect the social connection, and send us a message back if
they established the connection by mistake. As a result, we have sent out the message to 174 friends. Within 48 hours, 
we received 6 messages 
to clarify that the connection is mistakenly established. Also, we observed a significant decrease 
in 
the friends of the test accounts. 
In total, 16 friends out of 174 disconnect the connections to the fake accounts. 
This number is a conservative estimation of the real users that those fake accounts have connected to, because some real users
may have simply ignored our messages.

%
%
%
%
%

\subsection{Discussion} 
\label{subsec:discuss}
\paragraph{Non-manipulability of negative feedback.}
In our proposal, 
a user is able to signify a negative feedback only if she received unwanted 
communication such as an unexpected friend request, a spam message, or a spam post on her wall. 
This means that a negative feedback can be generated only if the user has been directly annoyed or harmed.
This is in contrast to the negative ratings in online services such as YouTube and Flickr, 
where users are granted to rate arbitrarily based on their preference.
In our proposal, 
a real user will not receive negative feedback if she never sends out unwanted communication.
We choose to use the abuse-triggered negative feedback because it is non-manipulable under collusion. Without the trigger of
abusive activities, 
a group of malicious users cannot collude to render arbitrarily negative feedback to a victim. 

\paragraph{Why not use negative feedback to directly detect Sybils?}\\
The negative feedback can be used to directly detect Sybils with machine-learning (ML) techniques.
However, ML-based techniques~\cite{Yang-IMC-11} 
require extensive calibration efforts due to the abundance of possible legitimate 
and malicious behaviors in OSNs. More importantly, these approaches are 
based on individual user 
features, and the resulting alarms or alerts are only applicable to individual users.
Therefore, such techniques may miss the Sybils behind the 
active entrance Sybils or currently silent entrance Sybils. 
Instead, \sys employs social-graph-based schemes and considers Sybils as groups. 
By aggregating negative feedback, \sys is able to leverage the 
aggressive behaviors of the entrance Sybils to uncover a much larger set of Sybils behind them.

%% file: system-design.tex
\section{system design}
\label{sec:design}
%
%
%


In the previous section, we discussed what is the user negative feedback and how would user negative feedback occur.
We now discuss how to incorporate user negative feedback into social-graph-based Sybil defenses.
We first introduce our system model and threat model.

\begin{figure}[t!]
\begin{center}
\epsfig{file=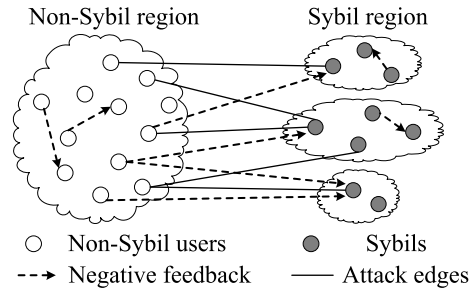, height=3.8cm}
\caption{\bf \footnotesize The social graph and the negative feedback graph in an OSN. 
\label{fig:system_model}}
\end{center}
\end{figure}

\paragraph{System model.} 
We model an OSN with two graphs:
1) the underlying social graph as an undirected graph $G^+ = (V, E^+)$, where $V$ is the user set in 
the OSN, and $E^+$ represents the social relationships among users;
and 2) the negative feedback graph as a directed graph $G^- = (V, E^-)$, 
where $V$ is the same user set as in $G^+$, $E^-$ is the directed edge set that includes the negative feedback between users. 
Given a direction between each pair of users, we consider no more than one negative feedback edge. 
Figure~\ref{fig:system_model} shows both the social graph and the negative feedback graph sharing the same 
user set in an OSN. 
A node $v$ has a {\it social degree} of $deg^+(v)$ in $G^+$ and an 
in-degree of $deg^-(v)$ in $G^-$.

\paragraph{Threat model.}
Malicious users may launch Sybil attacks by creating many fake accounts. We divide the 
user set $V$ into two disjoint subsets: non-Sybil users and Sybils, as shown in 
Figure~\ref{fig:system_model}. The {\it non-Sybil region} is the collection of the non-Sybil users, 
and the social edges and negative feedback links among them. Similarly, we define the {\it Sybil region} 
with respect to the Sybils.
We refer to the social edges between the 
non-Sybil region and the Sybil region as {\it attack edges}. We refer to the Sybils adjacent to the attack 
edges as {\it entrance Sybils}, and the other Sybils as {\it latent Sybils}. 
%
%

\subsection{Overview}

\sys aims to improve the social-graph-based Sybil defenses using the negative
feedback. 
Our observation is that the Sybils' attack edges are always accompanied by negative feedback.
If we discount the trustworthiness of the social edges that come with the negative feedback, 
we can limit the impact of the excessive attack edges on well-maintained entrance Sybils,
and enable social-graph-based Sybil defenses to uncover the Sybils behind them.

%
%

\sys comprises of two major modules: 
1) a {\it negative feedback combiner}, which incorporates the 
negative feedback graph into the social graph, and generates a {\it defense graph} with
discounted social edges on the users that have received negative feedbacks; and 
2) an {\it adopted social-graph-based defense scheme} 
that detects Sybils on the defense graph with improved accuracy.

%
%

\subsection{Incorporating negative feedback} 
\label{subsec:combining}

The social-graph-based Sybil defenses bound the accepted Sybil to the number of attack edges~\cite{Yu-SP-08, Cao-NSDI-12},
regardless of how much negative feedback that Sybils have received. 
\sys improves over social-graph-based Sybil defenses by reducing the impact of attack edge
through the collected user negative feedback.
In principle, we aim to build a weighted {\it defense graph} based on the social graph 
and the negative feedback graph.
We reduce the weights of social edges on the users that have received negative feedback,
such that the aggregate weight on attack edges is substantially limited.


\paragraph{Defense graph.}
Our key idea is to discount a user's social relationships with the
negative feedback that the user has received. By cancelling out the social edges that 
come along with negative feedback from other users, we are able to mitigate the impact of the
entrance Sybils' attack edges.
We define the {\it net social degree} of a node $v$ as $net(v) = deg^+(v) - \alpha \times deg^-(v)$ (we require $net(v) \ge 0$), 
where the offset factor $\alpha$ is a positive parameter.
A large $\alpha$ indicates a substantial penalty of an incoming negative feedback edge. 
Thus, a node whose social edges accompanied by negative feedback 
has a net social degree $net(v)$ 
smaller than its social degree $deg^+(v)$.

We define the weight of a node $v$ as its net social degree divided by its social degree: 
$w(v) = \frac{net(v)} {deg^+(v)}$. The node weight is essentially the discount rate of a node.
With this definition, a real user that never receives negative feedback 
has a weight of value $1$, while a user 
with received negative feedback is assigned
a discounted weight due to the discounted social degree. A node weight can be translated to
the extent to which the node can be trusted.

We derive the weight of a social edge based on the weights of its adjacent nodes:
$w(u, v) = \min(w(u), w(v))$. The weight of an edge is determined by 
the lowest weight of its adjacent nodes. This is because any of the 
adjacent nodes that has triggered negative feedback should discount the edge quality.
Therefore, an edge weight is always no more than $1$, and low weighted social edges are always
adjacent to incoming negative feedback links.

Therefore, we can build a weighted and undirected {\it defense graph} $G= (V, E^+, w)$. 
We weight each edge $(u, v)$ with $w(u, v)$ that discounts the quality
of social edges. 
As compared to the social graph, where every social edge is treated equally, 
our defense graph enforces strictly limited aggregate weight on attack edges using negative feedback.

\subsection{Detecting Sybils}
%
%
%

%
%
In our initial design, we take SybilRank~\cite{Cao-NSDI-12} as a proof of concept, and adapt it to the weighted defense graph. 
SybilRank comprises of three steps: trust propagation, trust normalization, and ranking. We adapt 
SybilRank to use the defense graph as below.

In the first stage, SybilRank propagates trust from the trust seeds via $O(\log |V|)$ power iterations.
In each iteration, the distributed trust on each edge is proportional to the edge weight. 
$T^{(i)}(v)$ is the amount of trust on node $v$ in the $i^{th}$ iteration.
We assume that the non-Sybil region part of the defense graph is well connected after social edges discounting.
We empirically validate it with simulations (\S\ref{sec:eval}), 
and leave a formal study 
in future work.  
In the second stage, SybilRank 
normalizes the total trust of every node with its social degree. Since a node's
social degree is always no less than the aggregate weight of its adjacent edges due to the discounted edge weights,
this normalization further penalizes Sybils that are likely to have substantially discounted social edges.
The last stage outputs a ranked list according to the degree-normalized trust with non-Sybil users on top.

\begin{small}
\begin{algorithm}[H]
\renewcommand{\thealgorithm} {}
\caption{$Adapted\_SybilRank (G(V, E^+, W), seedSet \text{ } S)$ 
\label{alg:srank}}
\begin{algorithmic}[r]
\STATE {\bf Stage I:} $O(\log |V|)$-step trust propagation 

\STATE In initialization, seed trust evenly in trust seeds;

\STATE In a step $i$ ($0 < i \le h$, $h = O(log |V|)$), node $u$ \\
updates its trust as below:

\STATE $T^{(i)}(u) = \sum_{(u,v) \in E^+} T^{(i-1)}(v){\frac {w(u,v)} {\sum_{(k,v) \in E^+} w(k, v)}}$

\STATE {\bf Stage II:} Normalize node trust by social degree

\quad \quad \quad \quad \quad \quad $\hat{T}_u = \frac {T^{(h)}_u} {deg^+(u)}$
 
\STATE {\bf Stage III:} Rank users based on their \\
degree-normalized trust $\hat{T}$

\STATE {\bf Return} ranked list $L$
\end{algorithmic}
\end{algorithm}
\end{small}

%% file: evaluation.tex
\section{Simulation study}
\label{sec:eval}

To gain a better understanding on how our initial design improves the detection accuracy,
we evaluate \sys 
in comparison to the original SybilRank. 
We simulate user friend requests on social graphs, and use the request rejections
as negative feedback. 

%
%

%
%
%
%
%
%
%

\subsection{Simulation setup}
We simulate Sybil attacks in four social graphs (Table~\ref{table:dataset}). 
The Facebook graph is sampled via the ``forest fire'' sampling method~\cite{Leskovec-KDD-06}.
The synthetic graph is generated based on the scale-free model~\cite{Barabasi-Science-99}.
We connect a Sybil region that consists of 5,000 Sybils to each social graph. 
We simulate the social connections among Sybils by establishing 
social edges from each Sybils to another 5 random Sybils upon its arrival.

%
%

\begin{table}[ht!]
\begin{center}
\begin{footnotesize}
\begin{tabular}{c|c|c|c|c} 

\textbf{Social} 	    & \textbf{Nodes}    & \textbf{Edges}  & \textbf{Clustering} & \textbf{Diameter} \\    
\textbf{Network}	    & 			& 		  & \textbf{Coefficient}  &  \\   \hline \hline   
\textbf{Facebook}           & $10,000$          & $40,013$    	  & $0.2332$  & $17$ \\     
\textbf{ca-AstroPh}~\cite{stanford_data}         & $18,772$          & $198,080$       & $0.3158$  & $14$ \\          
\textbf{ca-HepTh}~\cite{stanford_data}          & $9,877$           & $25,985$    	  & $0.2734$  & $18$ \\
\textbf{Synthetic}	    & $10,000$		& $39,399$	  & $0.0018$  & $7$ \\ 
     			
\end{tabular}
\end{footnotesize}
\end{center}
\caption{\footnotesize{Social graphs used in our simulation.}}
\label{table:dataset}
\end{table}

Similar to~\cite{Cao-NSDI-12} and~\cite{Viswanath-SIGCOMM-10}, we use the metric {\it the area under
the Receiver Operating Characteristic (ROC) curve}~\cite{Hanley-Radiology-1982} to compare the 
quality of the ranking that social-graph-based schemes use to uncover Sybils. 
The area under the ROC curve measures the probability that a Sybil user is ranked
lower than a random non-Sybil user. It ranges from 0 to 1. 


%
%

%
%

%
%
%

\subsection{Simulating negative feedback}
\label{subsec:simulating_feedback}
Users can send friend requests to 
others. 
%
%
A request acceptance yields a social edge, while a rejection produces a negative feedback edge.

\paragraph{Rejections to Sybil users.}
We simulate the process that the entrance Sybils solicit social edges from non-Sybil users. 
In the Sybil region, we designate 200 nodes as entrance Sybils, and the 
rest 4800 nodes as latent Sybils. The entrance Sybils represent well-maintained 
fake accounts, 
which continuously send friend requests and have lower rejection rates
than latent Sybils due to the better maintenance.
By default we set the rejection rate of entrance Sybils by non-Sybil users 
to 60\%, and the rejection rate of latent Sybils to 98\%.

\paragraph{Rejections to non-Sybil users.}
We simulate rejections to non-Sybil users based on the social graph.
In particular, given a rejection rate to non-Sybil users and the number of friends
that a non-Sybil user has in the social graph, we can infer the number of rejections 
on this user. 
we then add this number of rejections to the non-Sybil user by randomly selecting
non-friend users and simulating a rejection from each of them.
We set the rejection rate of non-Sybil users to 1\%. 
We study how \sys's performance
varies with the change of the rejection rates in \S\ref{subsec:results}. 

%
%

\subsection{Simulation results}
\label{subsec:results}
\ifthenelse{\boolean{TECHREPORT}} { 
We now present the simulation results in the Facebook graph. The results on other graphs are similar (see Appendix).
}
{ 
Due to the limited space, we only present the simulation results in the Facebook graph, and refer the reader
to our technical report for the complete results~\cite{Cao-Rejection-12}.
}

\begin{figure}[t!]
\begin{center}
\epsfig{file=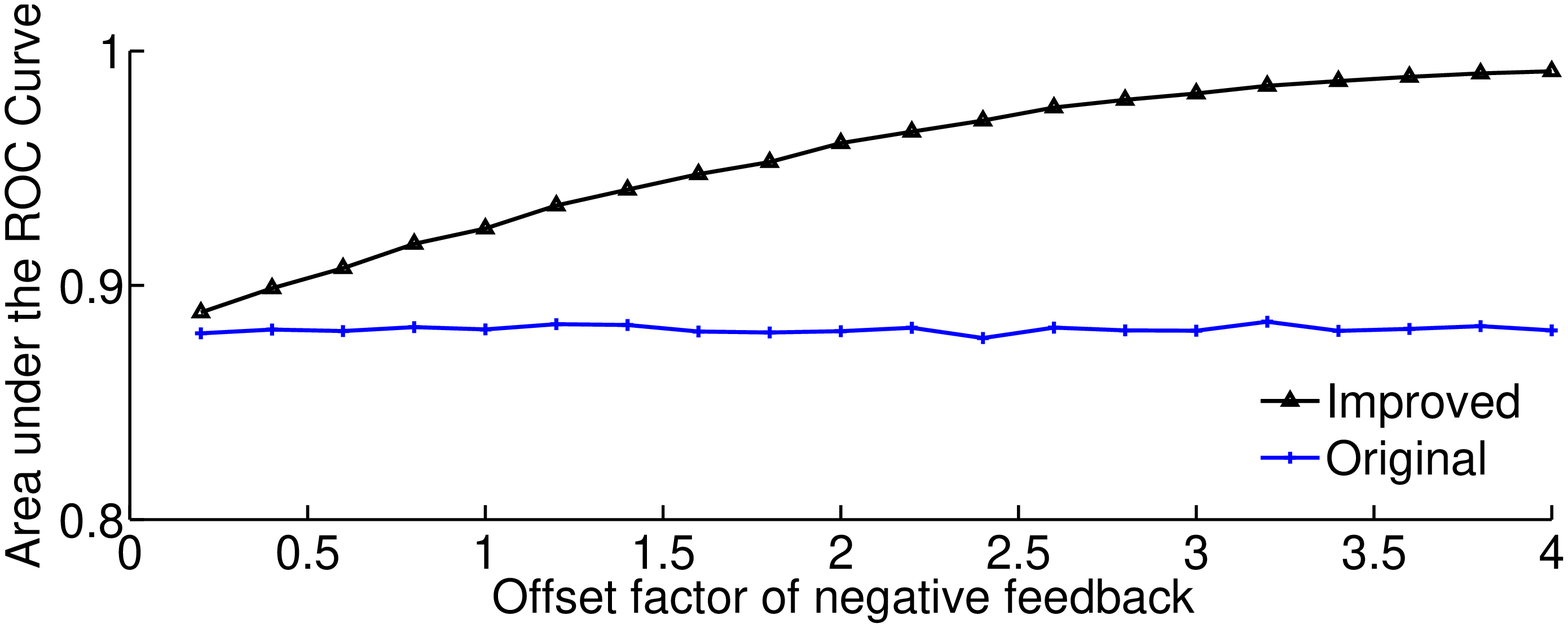, height=3.1cm}
\caption{\bf \footnotesize Impact of the offset factor of negative feedback.}
\label{fig:penalty}
\end{center}
\end{figure}

\paragraph{Impact of the negative-feedback offset factor.}
The offset factor $\alpha$ (\S\ref{subsec:combining}) is a penalty factor to the nodes that have received rejections,
including both Sybils and non-Sybil users. To investigate its impact, we vary the value of the penalty factor from 0 to 4.
Since we set the rejection rate of entrance Sybils' requests to 60\%, 
an offset factor of 
value $\frac {2} {3}$ can leverage the rejections to cancel out 
the attack edges on entrance Sybils.
However, the entrance Sybils also have social edges from Sybils. 
Therefore, a larger offset factor can yield further improvement.
Figure~\ref{fig:penalty} shows that the improvement keep increases 
until the offset factor reaches a sufficient large value, i.e., 3.0. With this value, \sys is able to cancel
out most of the entrance Sybils' social edges from both non-Sybil users and Sybils.

%
%

\begin{figure}[t!]
\begin{center}
\epsfig{file=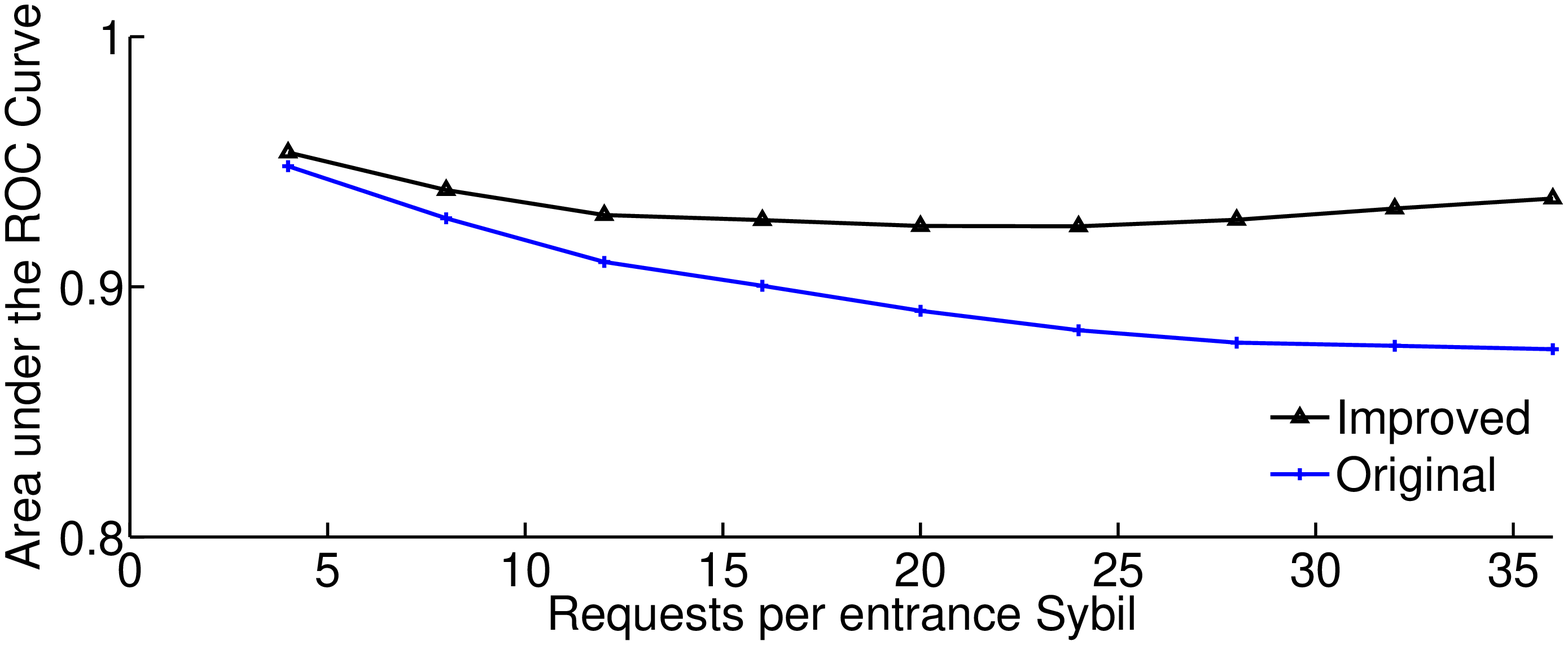, height=3.1cm}
\caption{\bf \footnotesize Resilience to the number of requests from entrance Sybils.}
\label{fig:probe}
\end{center}
\end{figure}

\paragraph{Resilience to Sybils' flooding requests.}
The fake accounts can solicit social edges by flooding friend requests.
As a result, the attack edges keep increasing.
We study the \sys's resilience to request flooding. We set the offset factor to 1 and
vary the number of requests that each entrance Sybil sends from 4 to 36. 
Each latent Sybil is set to send 2 requests to random non-Sybil users. 
Consequently, the attack edges increase from $\sim$500 to $\sim$3000. Figure~\ref{fig:probe} shows 
that \sys has only small performance degradation. 
In contrast, SybilRank's performance decreases sharply, because its security guarantee only relies on the 
number of attack edges.

%
%

\begin{figure}[t!]
\begin{center}
\epsfig{file=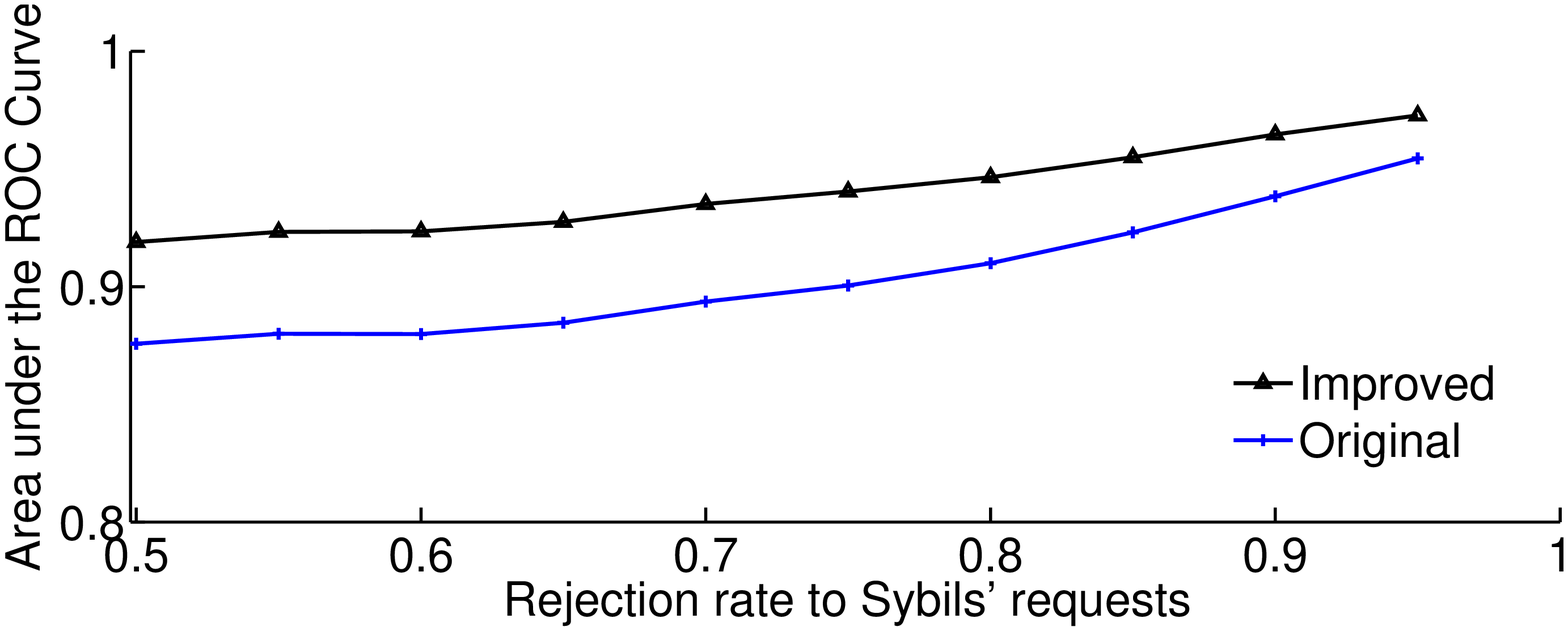, height=3.1cm}
\caption{\bf \footnotesize Detection accuracy as a function of the rejection rate to the Sybils' requests.}
\label{fig:sybilrej}
\end{center}
\end{figure}

\paragraph{Impact of the rejection rate to Sybils' requests.}
At a high level, both \sys and social-graph-based schemes rely on non-Sybil users to defense against
Sybils.
%
%
We investigate the impact of the rejection rate to Sybils' requests.
In this simulation, each entrance Sybil sends 25 requests to random non-Sybil users.
We vary the rejection rate to these requests from 0.5 to 0.95. The number of attack edges decreases accordingly
from $\sim$2700 to $\sim$450.
An increased rejection rate to 
Sybils' requests improves both \sys and SybilRank as shown in Figure~\ref{fig:sybilrej}, because this further 
limits the attack edges and signifies more negative feedback. \sys achieves higher accuracy due to its advantage
from the consideration of negative feedback.

%
%

\begin{figure}[t!]
\begin{center}
\epsfig{file=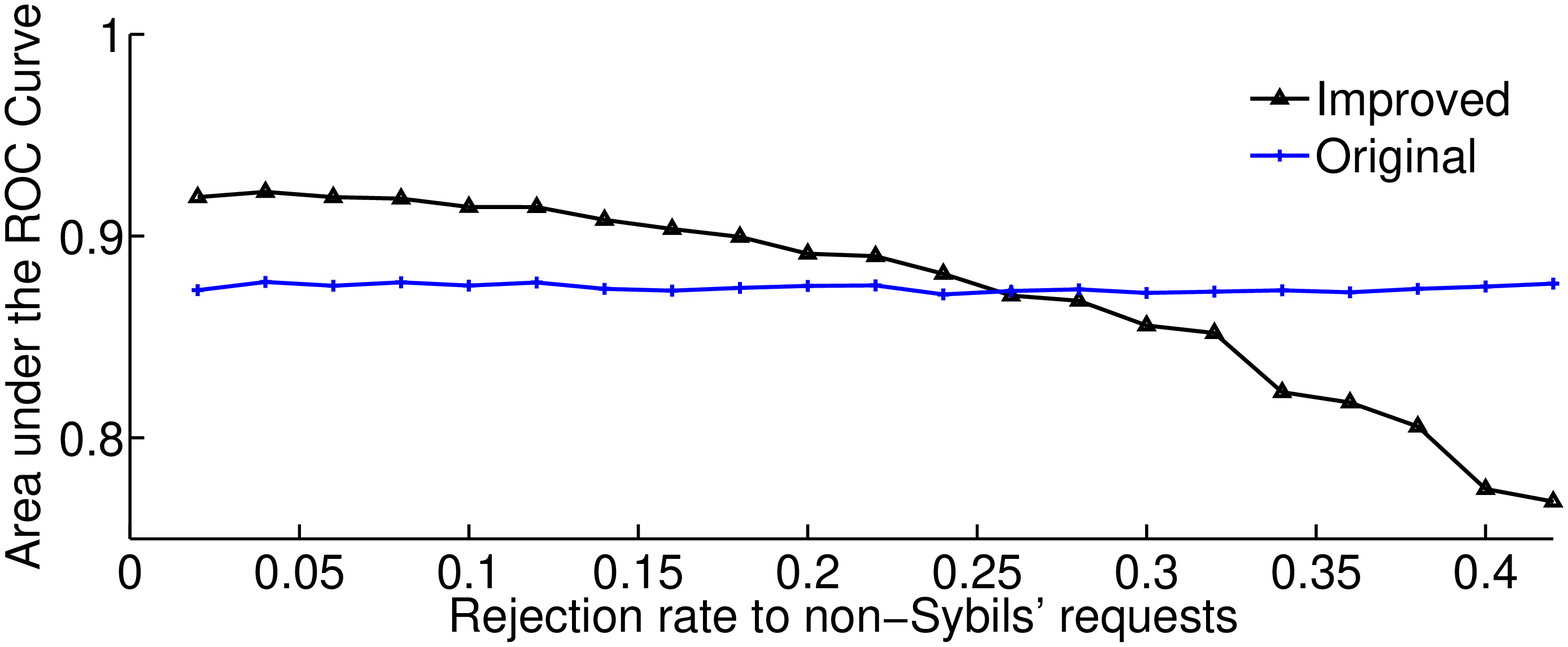, height=3.1cm}
\caption{\bf \footnotesize Detection accuracy as a function of the rejection rate to non-Sybil users' requests.}
\label{fig:nonsybilrej}
\end{center}
\end{figure}

\paragraph{Rejections among non-Sybil users.}
\sys is based on an assumption that non-Sybil users are less likely to receive negative back from others than
Sybils. 
We study \sys's performance with a varying rejection rate to non-Sybils' requests.
We simulate the rejections among non-Sybil users as decribed in \S\ref{subsec:simulating_feedback}.
The rejection rate increases from 0.05 to 0.45. The offset factor is set to 1 and the rejection rate
to entrance Sybils' requests is set to 0.4. 
As shown in figure~\ref{fig:nonsybilrej}, the improvement that \sys makes over SybilRank 
decreases as the rejection rate to non-Sybil users' requests increases. This is because more rejections to non-Sybil
bring more penalty to non-Sybil users, and thus increase the chance that a non-Sybil user ranks low in \sys.
\sys performs even worse than SybilRank when the rejection rate to non-Sybil users reaches 0.25. 
The reason is that beyond this rejection rate, the non-Sybil users are more likely to get rejections than
Sybils. This threshold is smaller than the entrance Sybils' rejection rate 0.4, because the entrance Sybils 
have 0 rejection rate to establish social edges among themselves. 
This results indicate that \sys cannot improves SybilRank if non-Sybil users are 
more likely to be rejected. We suspect that this does not happen in real world, because real users always 
send friend requests to acquaintances. We leave a further study to our future work. 

%
%

%% file: related-work.tex
\section{related work}
\label{sec:related}

This work is mainly related to social-graph-based Sybil defenses~\cite{
Yu-SP-08, Danezis-NDSS-09, Tran-NSDI-09, Viswanath-SIGCOMM-10, Cao-NSDI-12}. These
proposals rely on social graph properties to distinguish Sybils from non-Sybil users,
i.e., 
the attack edges are strictly limited. Existing schemes bound the accepted 
Sybils to the number of attack edges. Thus, fake accounts benefit from 
soliciting social connections.
\sys improves over existing Sybil defenses by leveraging negative feedback 
from users. It can be used to further uncover the fake accounts that 
have obtained social edges to real users but inevitably received negative 
feedback from resistant users.

There have been proposals to propagate distrust 
in social graphs~\cite{Guha-WWW-04, Borgs-WINE-10, Kunegis-WWW-09, Kerchove-SDM-08, Ziegler-ISF-05}.
However, those approaches are proposed as general techniques for social network analysis, but not 
targeting Sybil defense. Furthermore, the distrust in those proposals is not securely defined, 
which can include arbitrarily negative information and is not resilient to user collusion.

%% file: conclusion.tex
\section{conclusion and future work}
\label{sec:conclusion}

The detection of fake accounts in OSNs has been increasingly urgent
as both OSN operators and users have been suffering from 
illegal exploitation. 
We observe that even well-maintained fake accounts inevitably
receive negative feedback from others, 
as the controllers only have 
limited knowledge on users' security awareness.
Thus, user negative feedback can be used to strengthen existing Sybil defenses.
%
%
We propose \sys, which incorporates user negative feedback into social-graph-based Sybil defenses. 
Fake accounts can evade \sys 
only if they can connect to real users, 
and meanwhile receive little 
negative feedback from others.
Therefore, \sys advances the Sybil defenses by raising the cost for Sybils to evade detection.

In future work, we plan to continue our study on the live fake accounts in black market 
to further quantify the negative feedback they receive. With this study, we plan to complete and extend 
the our preliminary \sys design. We can then implement the complete \sys 
system and probably deploy it in real OSNs.

%% file: appendix.tex
\ifthenelse{\boolean{TECHREPORT}}
{ 
\appendix
\label{sec:appendix}

Figure \ref{fig:results-ca-AstroPh}, Figure \ref{fig:results-ca-Hepth}, and Figure \ref{fig:results-Syn}
show the simulation results in the graphs ca-AstroPh and ca-HepTh, and 
the synthetic graph (Table \ref{table:dataset}). In these graphs, SybilFence
achieves similar improvement over SybilRank as described in \S\ref{subsec:results}.

\begin{figure}[t!]
\centering 
\subfigure[Offset factor]{\label{fig:ca-astroph-penalty-factor}
  \includegraphics[scale=0.15]{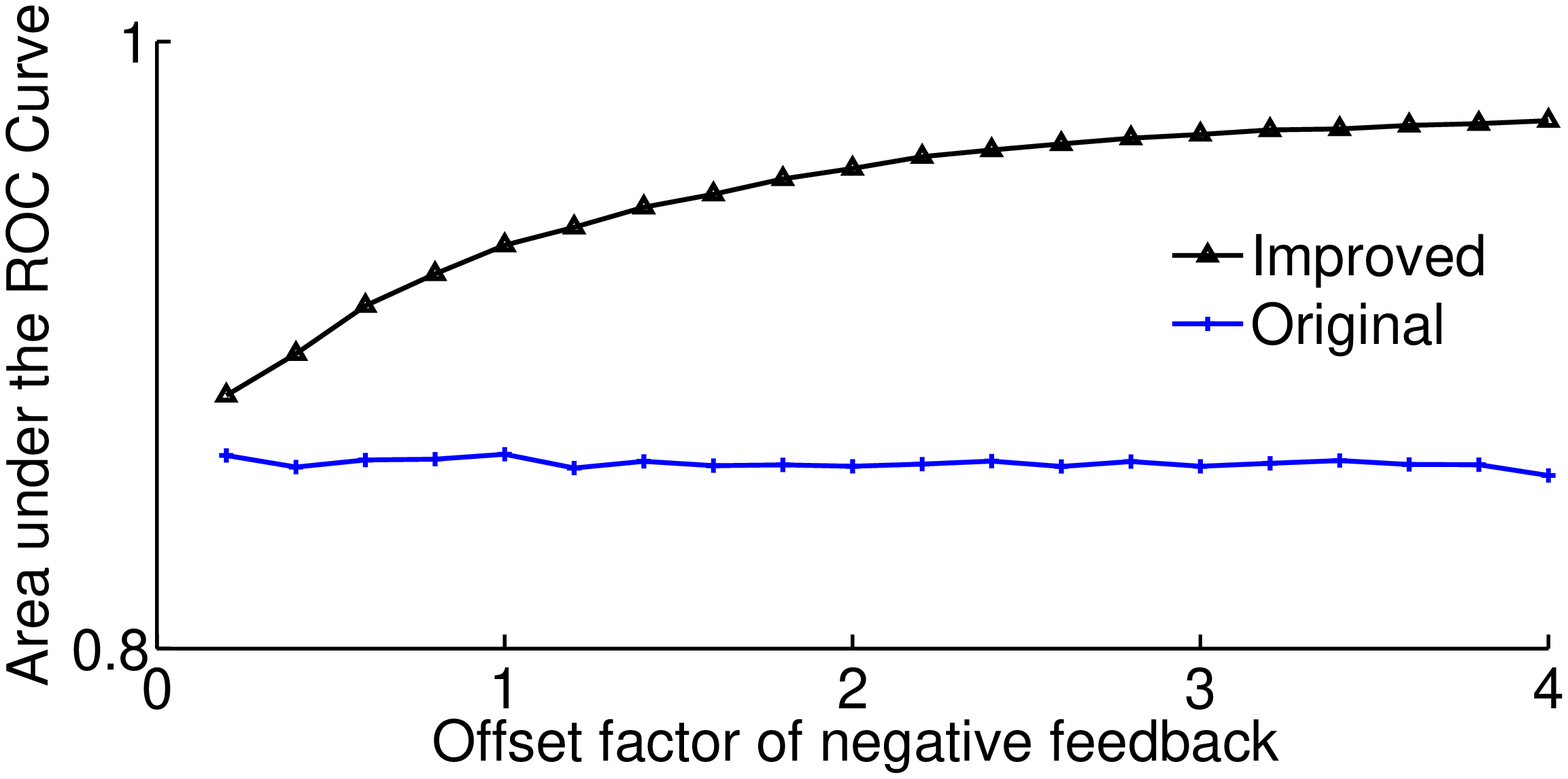}}
\subfigure[Flooding requests]{\label{fig:ca-astroph-probes}
  \includegraphics[scale=0.15]{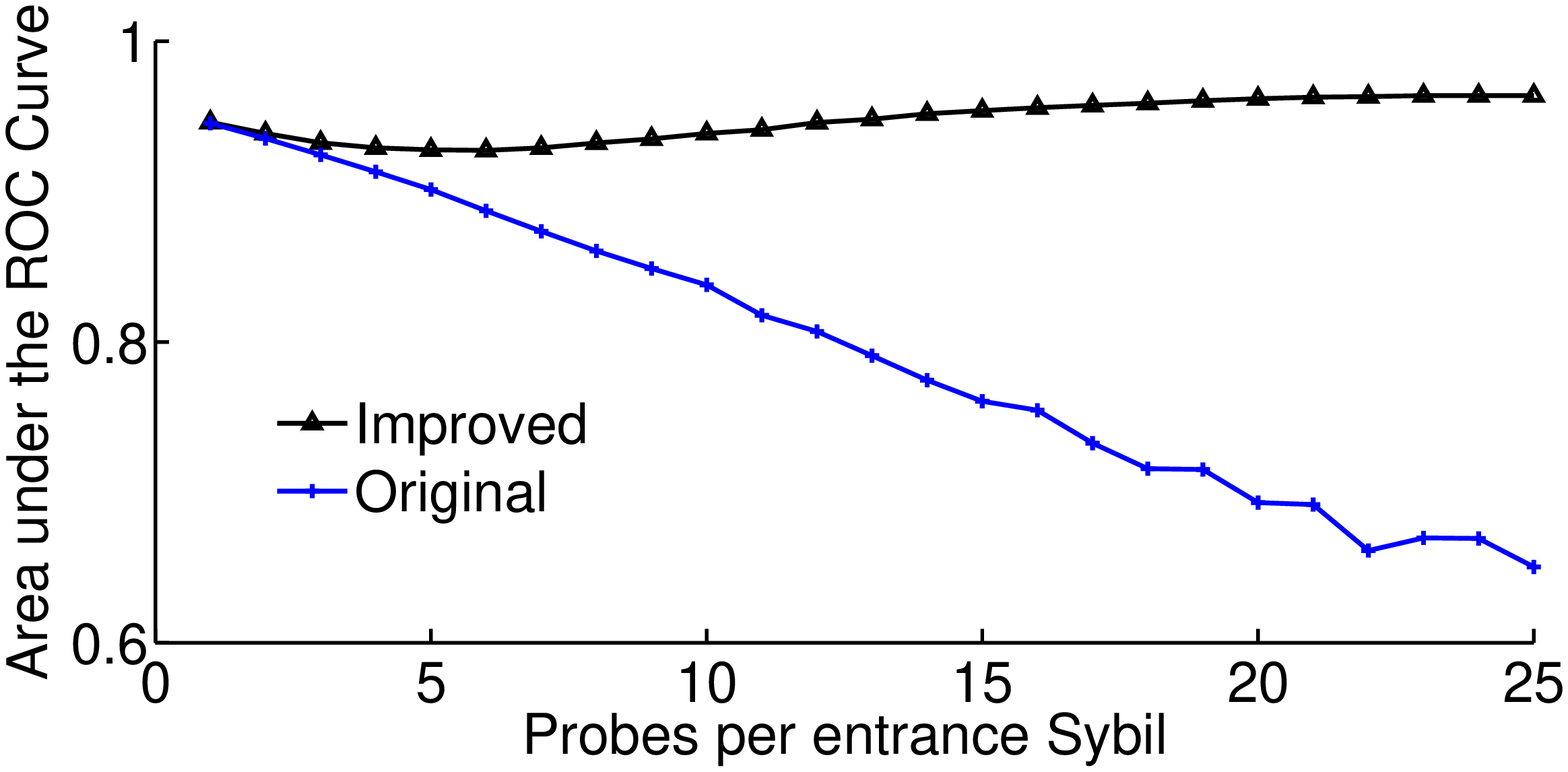}}
\subfigure[Rejection rate to Sybil requests]{\label{fig:ca-astroph-sybilrej}
  \includegraphics[scale=0.15]{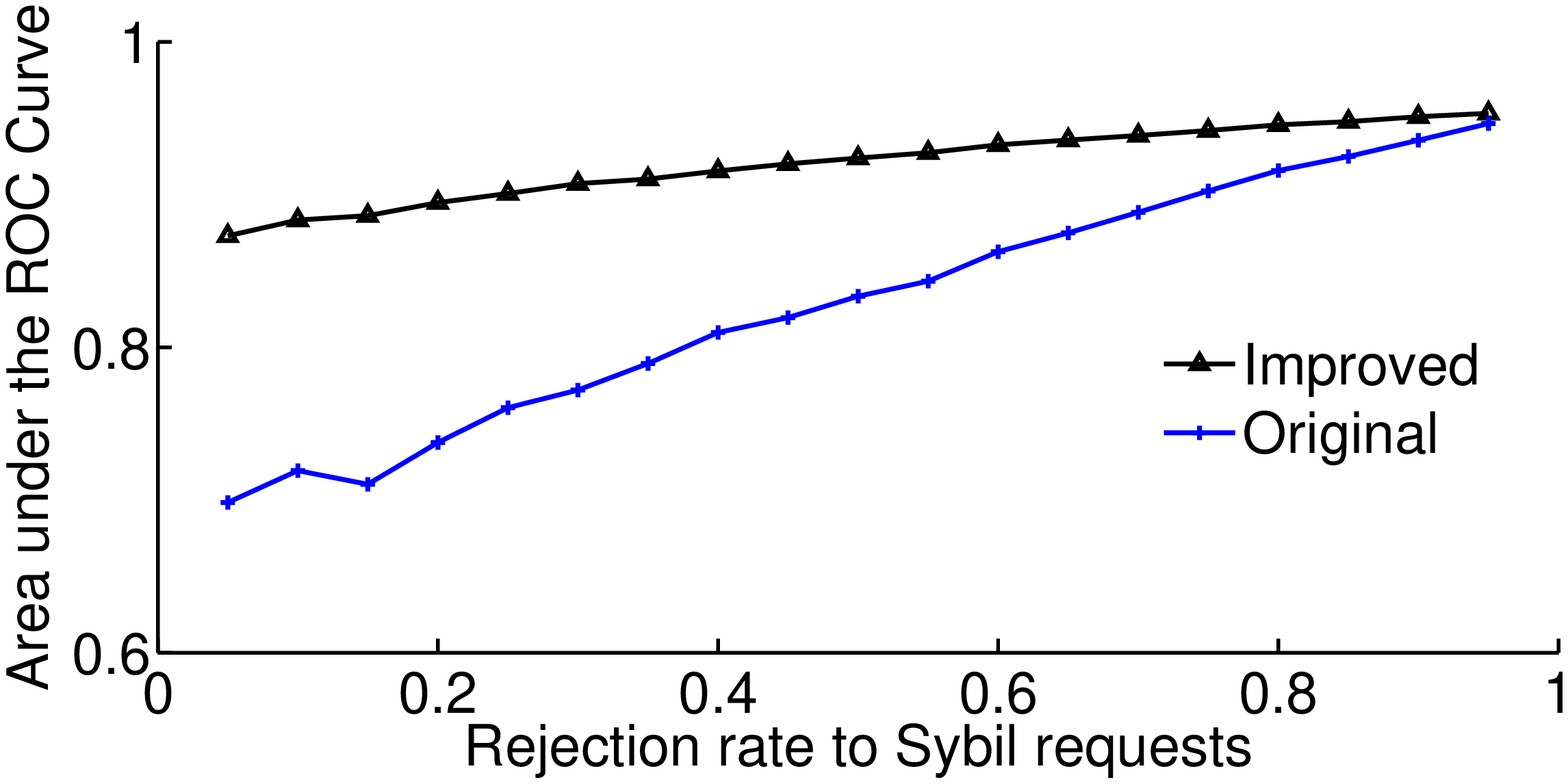}}
\subfigure[Rejection to non-Sybil requests]{\label{fig:ca-astroph-nonsybilrej}
  \includegraphics[scale=0.15]{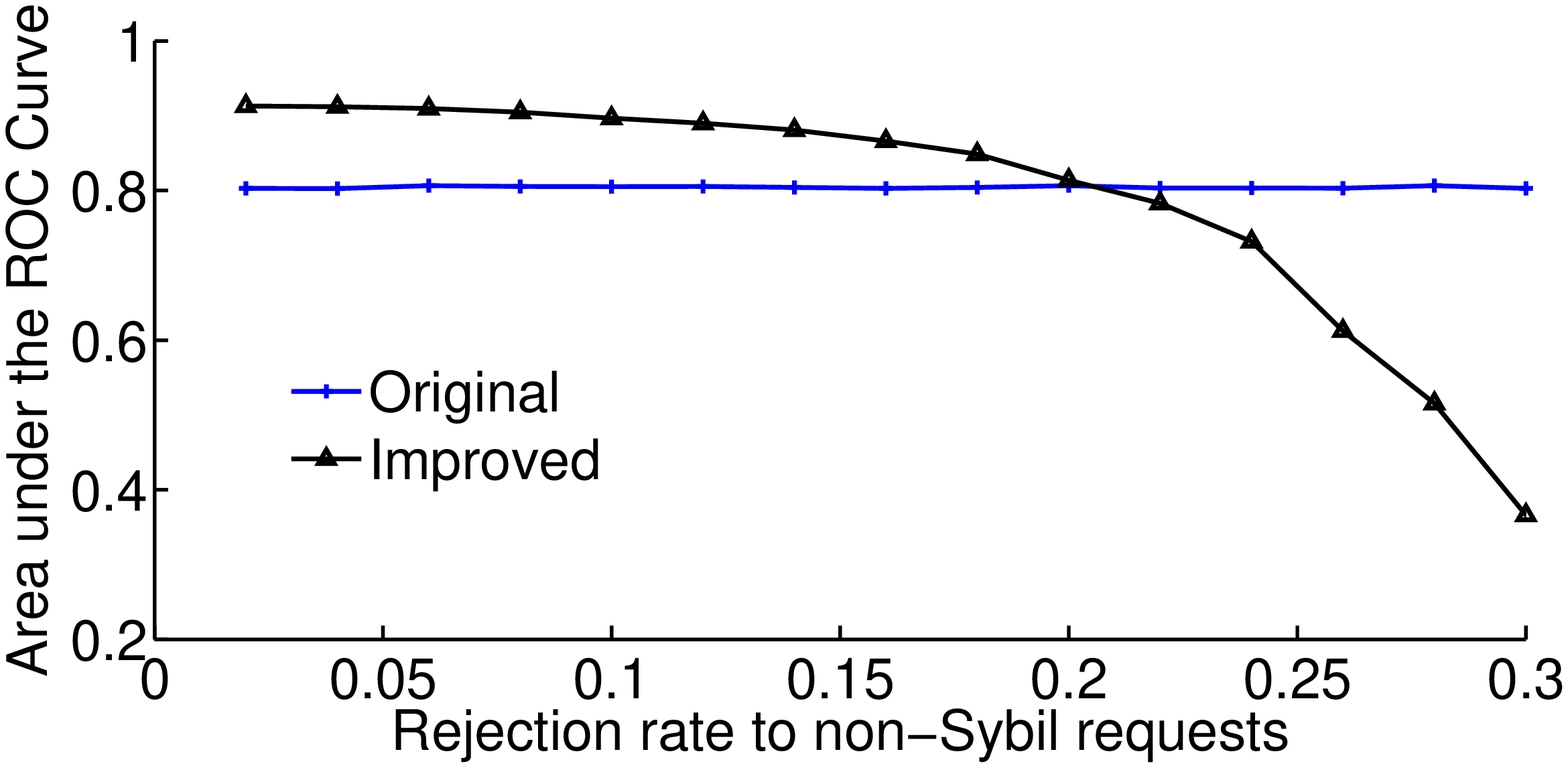}}
\caption{{\bf \footnotesize Simulation results in ca-AstroPh.
\label{fig:results-ca-AstroPh}}}
\end{figure}

\begin{figure}[t!]
\centering 
\subfigure[Offset factor]{\label{fig:ca-HepTh-penalty-factor}
  \includegraphics[scale=0.15]{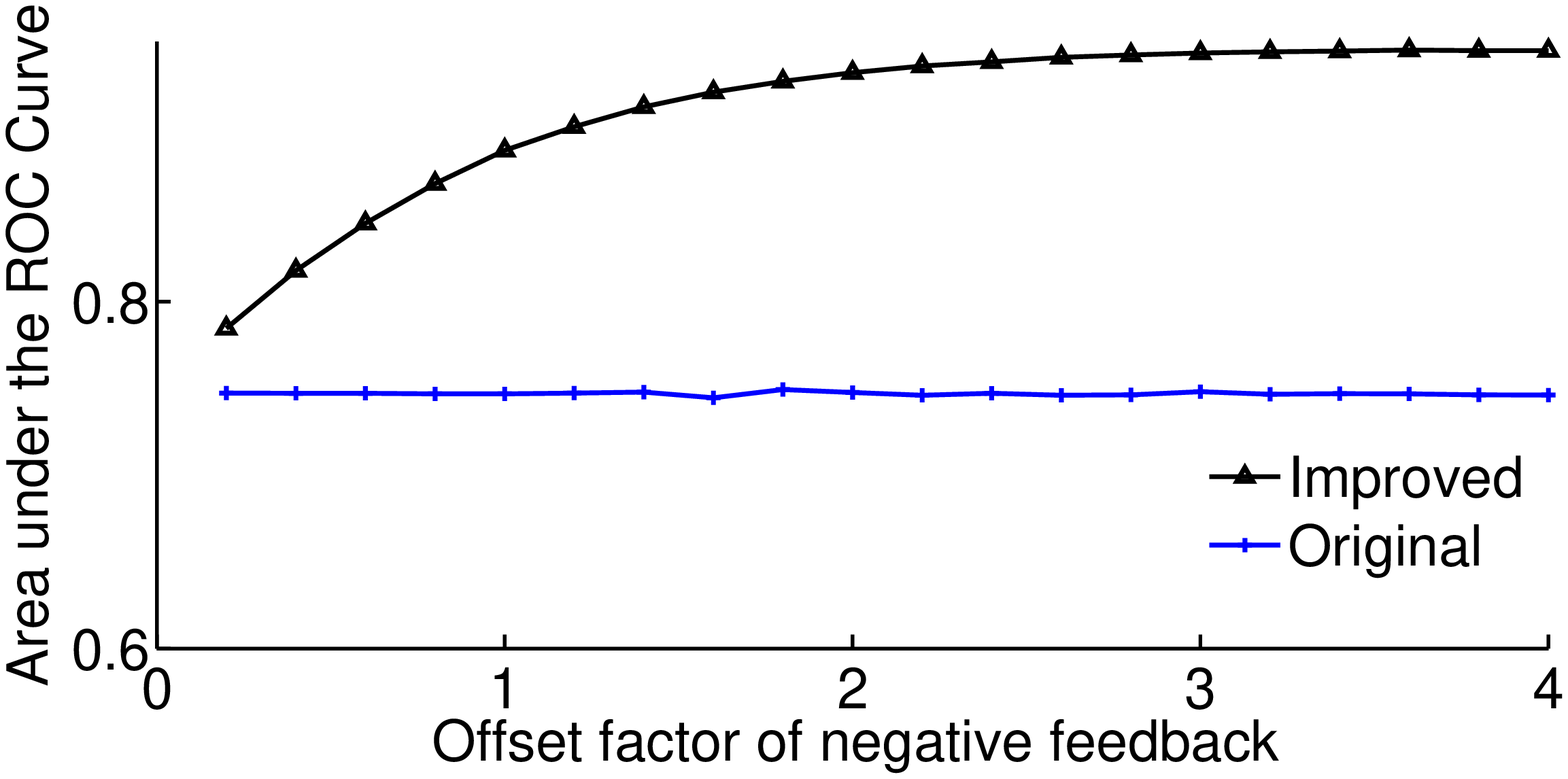}}
\subfigure[Flooding requests]{\label{fig:ca-HepTh-probes}
  \includegraphics[scale=0.15]{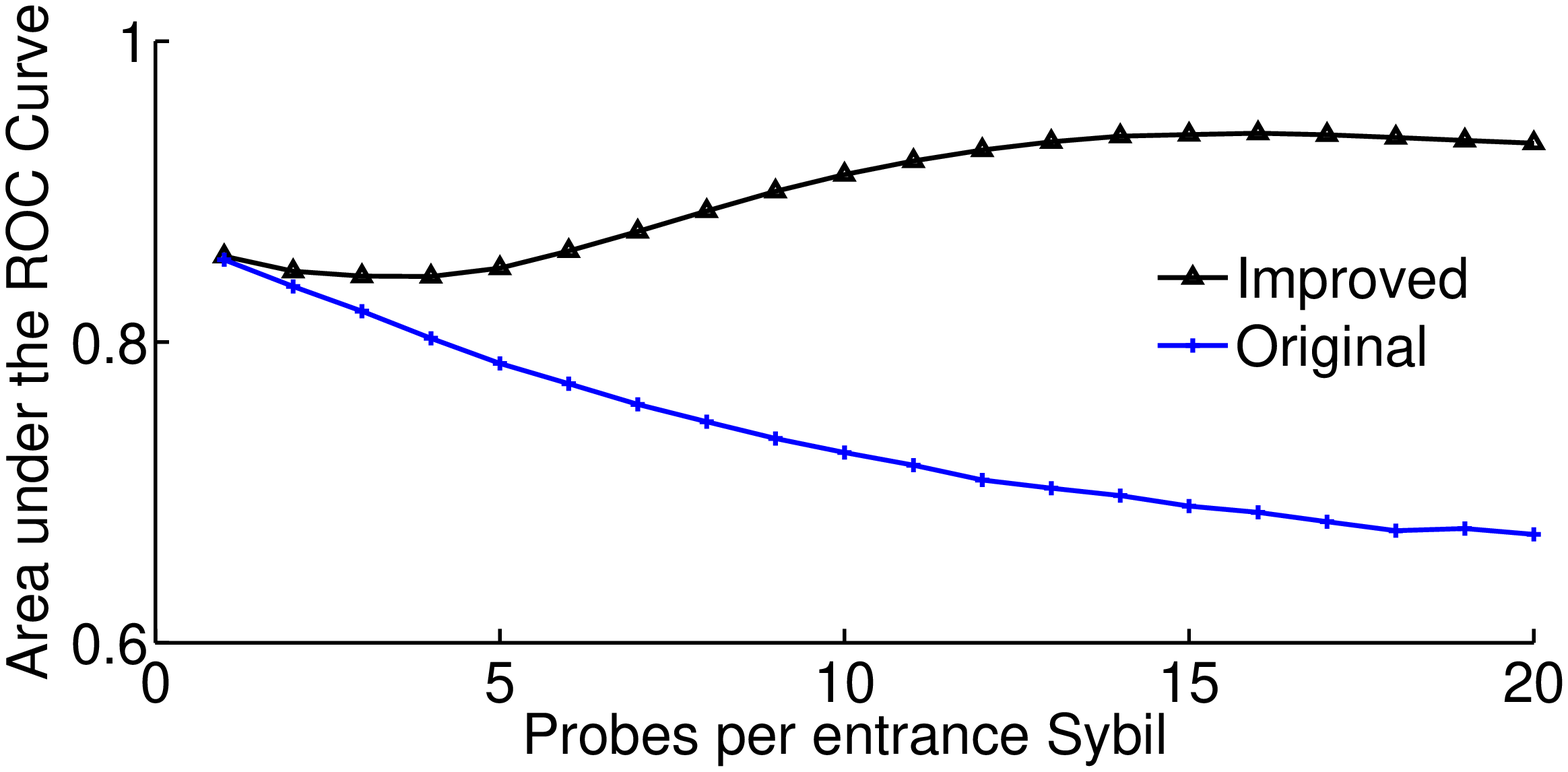}}
\subfigure[Rejection rate to Sybil requests]{\label{fig:ca-HepTh-sybilrej}
  \includegraphics[scale=0.15]{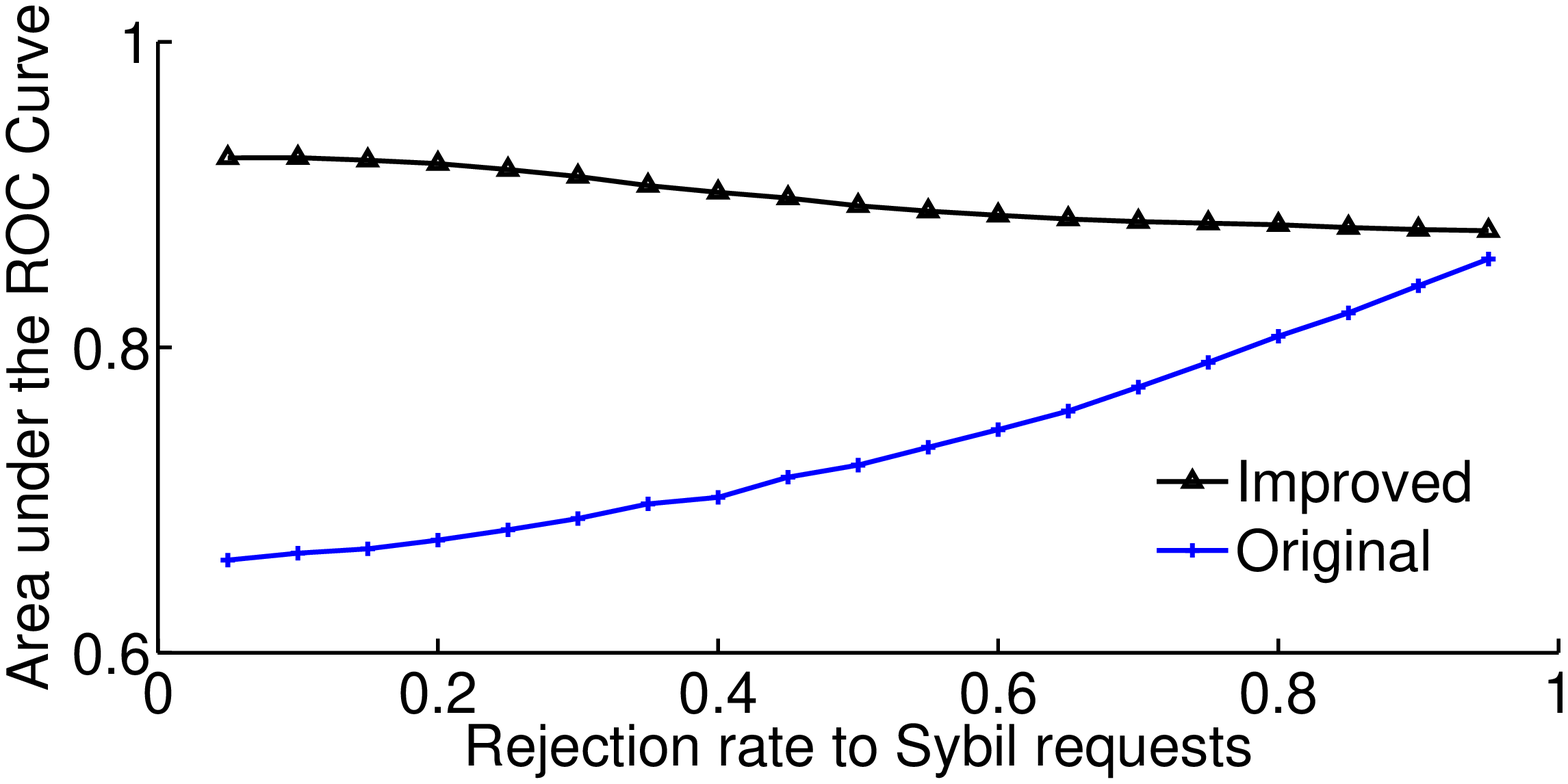}}
\subfigure[Rejection rate to non-Sybil requests]{\label{fig:ca-HepTh-nonsybilrej}
  \includegraphics[scale=0.15]{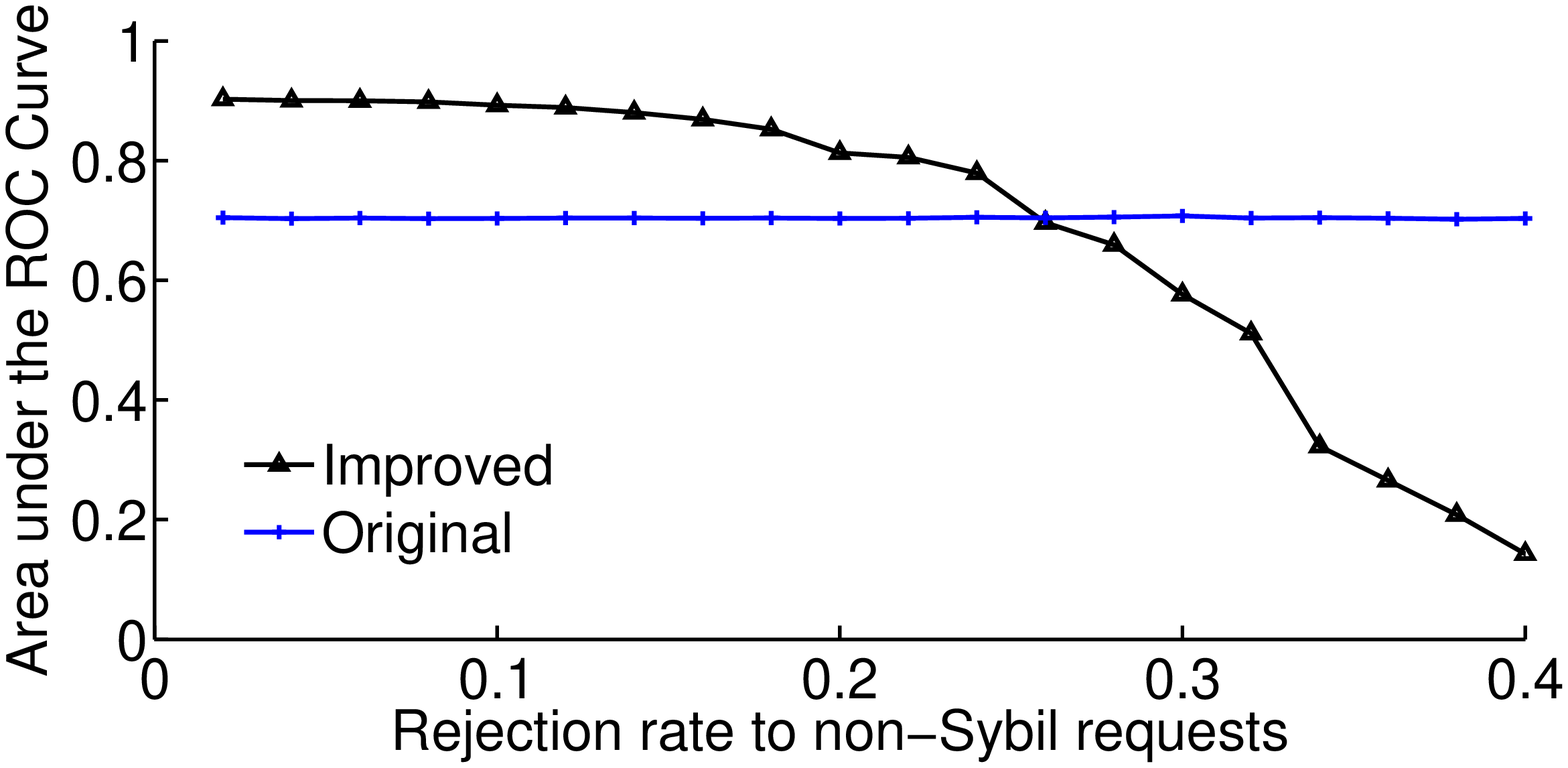}}
\caption{{\bf \footnotesize Simulation results in ca-HepTh.
\label{fig:results-ca-Hepth}}}
\end{figure}

\begin{figure}[t!]
\centering 
\subfigure[Offset factor]{\label{fig:syn-penalty-factor}
  \includegraphics[scale=0.15]{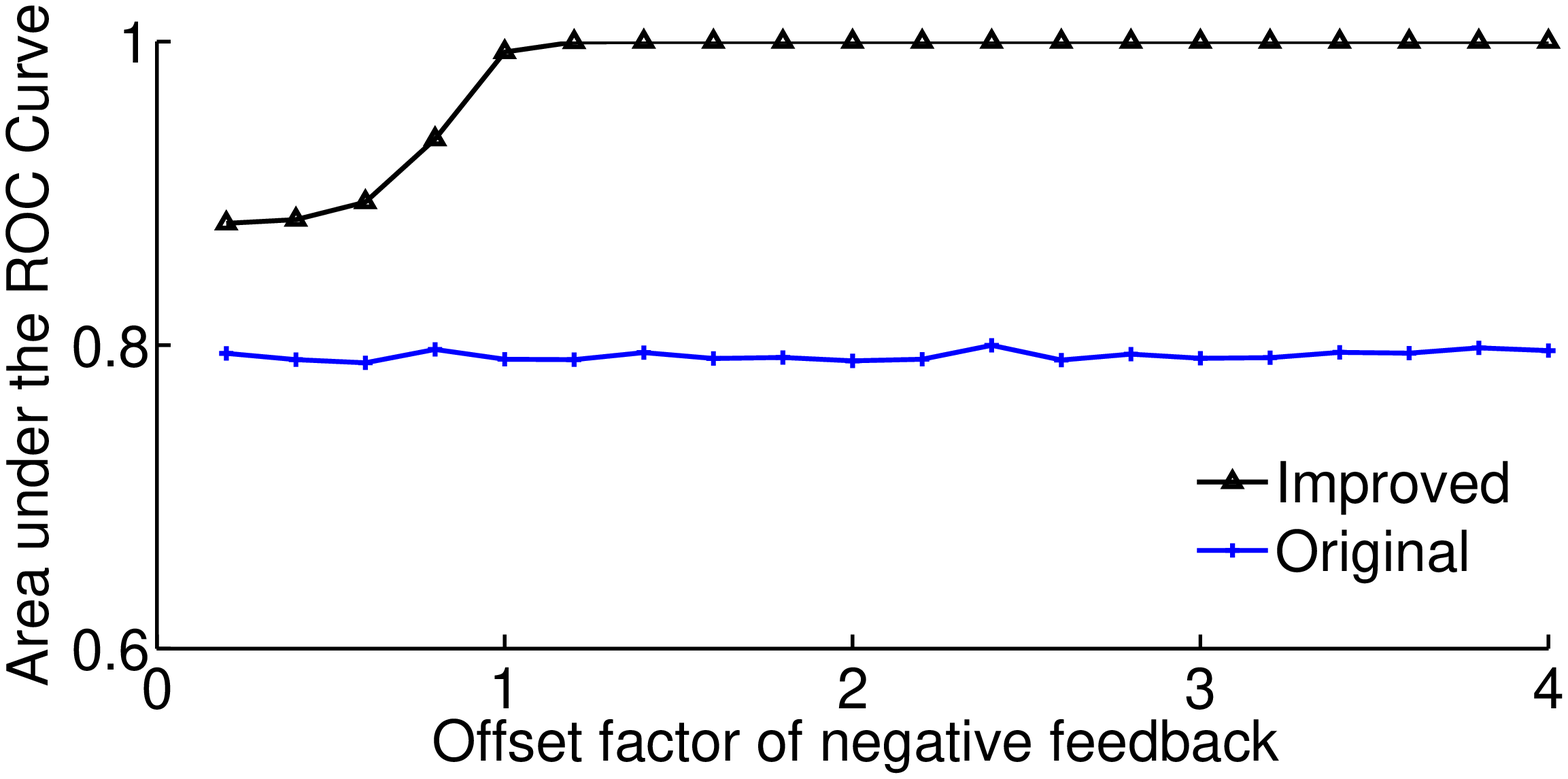}}
\subfigure[Flooding requests]{\label{fig:syn-probes}
  \includegraphics[scale=0.15]{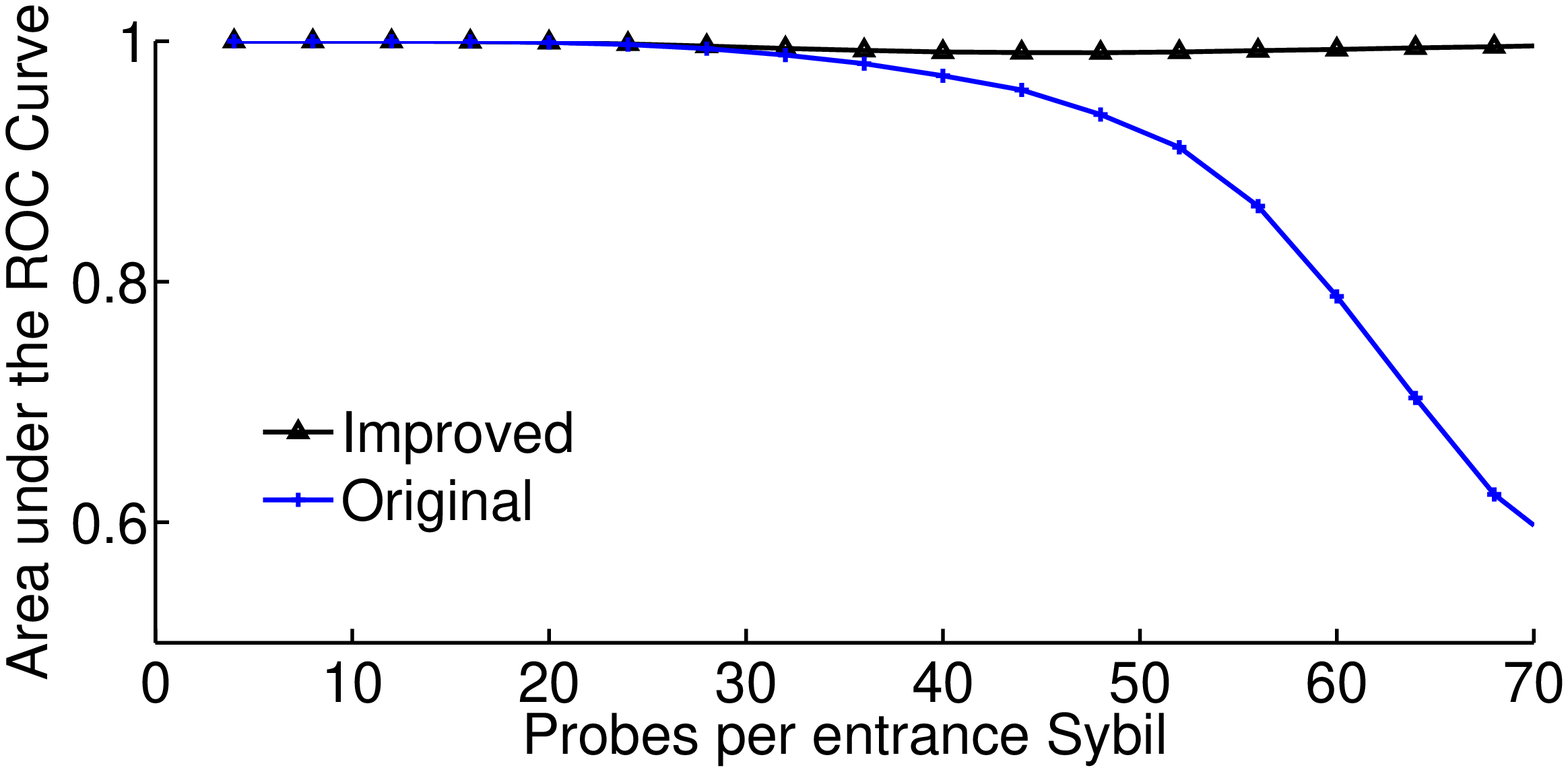}}
\subfigure[Rejection rate to Sybil requests]{\label{fig:syn-sybilrej}
  \includegraphics[scale=0.15]{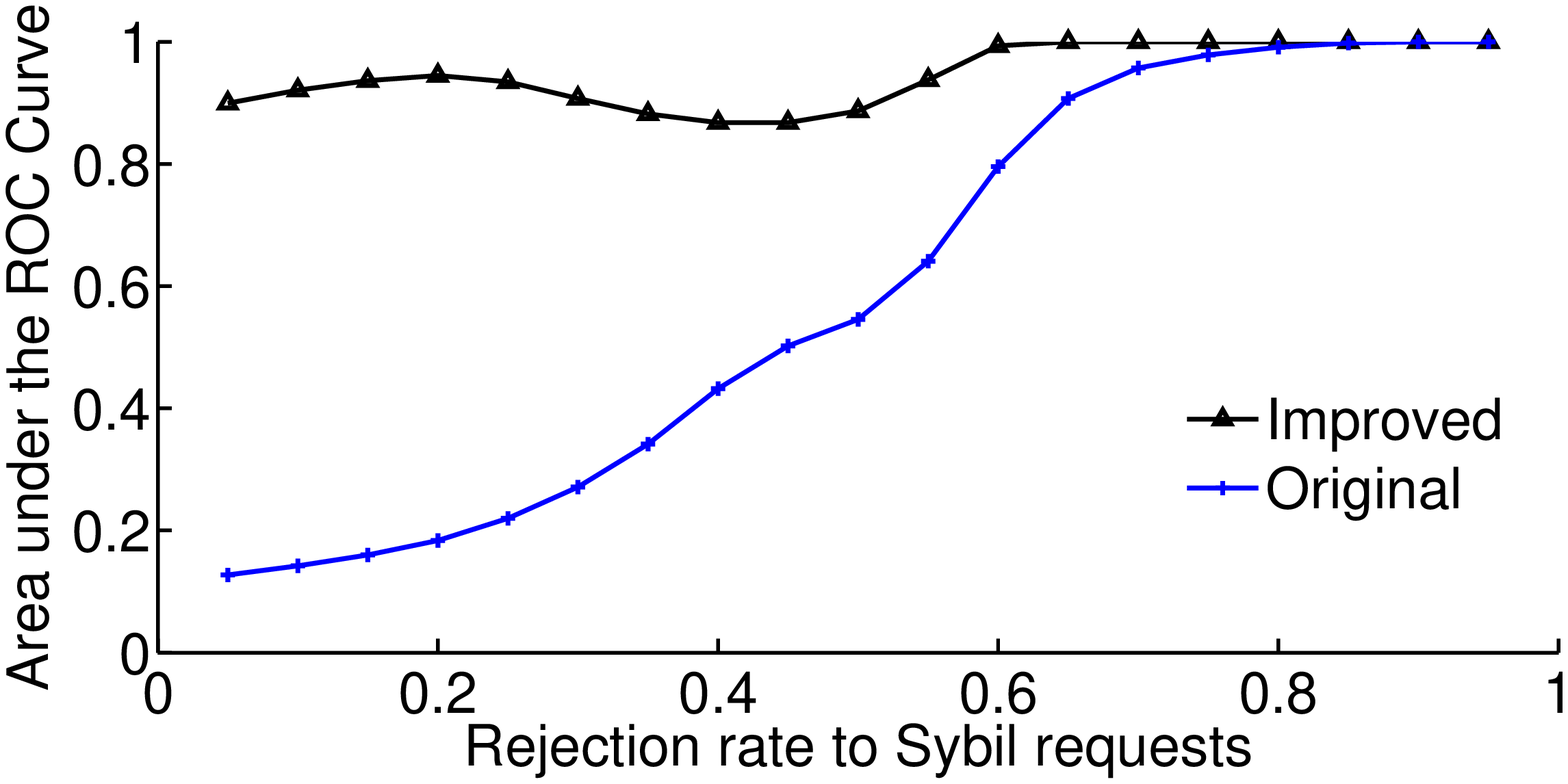}}
\subfigure[Rejection rate to non-Sybil requests]{\label{fig:syn-nonsybilrej}
  \includegraphics[scale=0.15]{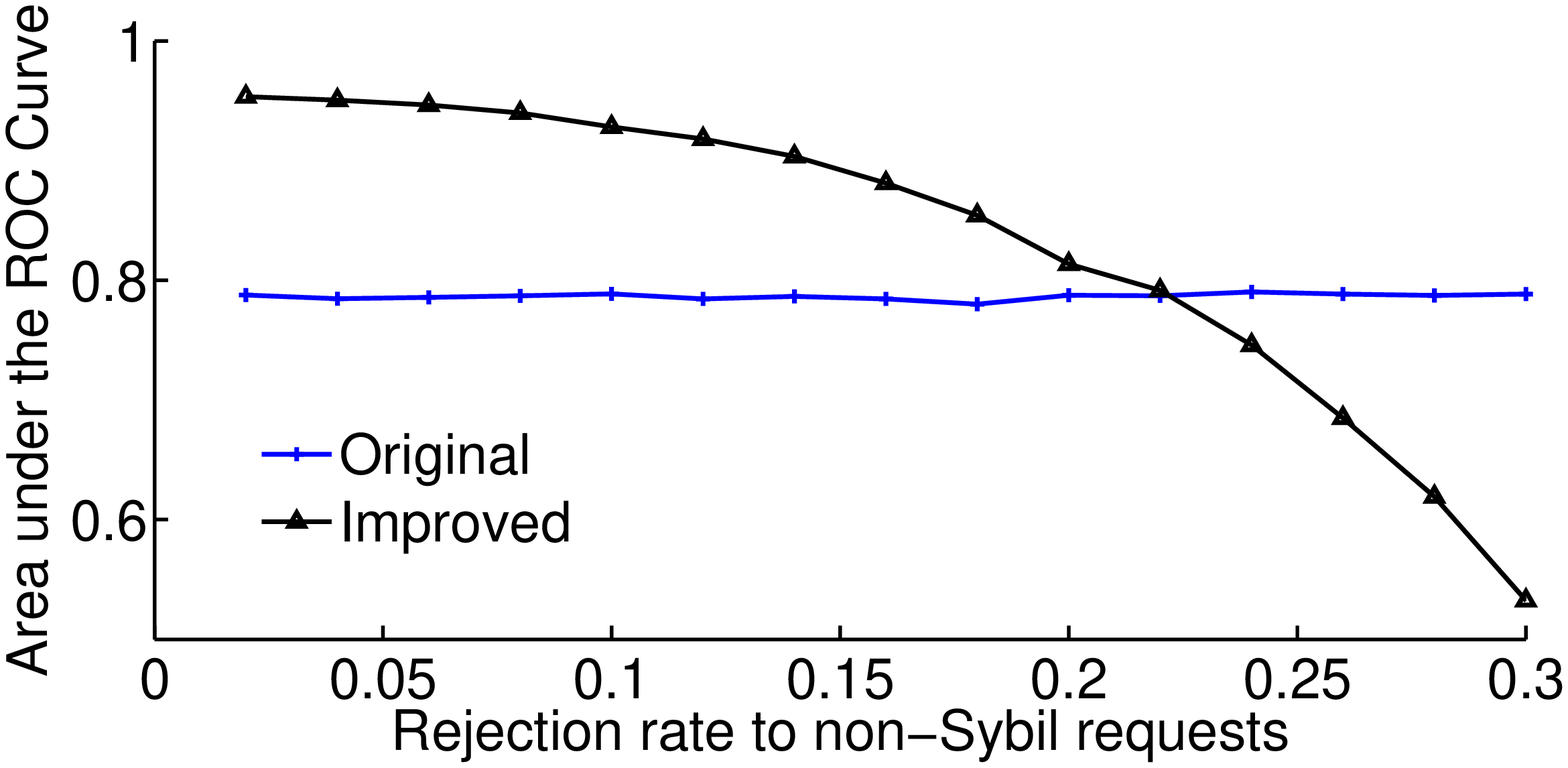}}
\caption{{\bf \footnotesize Simulation results in the synthetic graph.
\label{fig:results-Syn}}}
\end{figure}

}  

{  

}

%% file: paper.bbl
\begin{thebibliography}{10}

\bibitem{BlackHatWorld}
{BlackHatWorld}.
\newblock \url{http://www.blackhatworld.com/}.

\bibitem{Freelancer}
{Freelancer}.
\newblock \url{http://www.freelancer.com/}.

\bibitem{stanford_data}
{Stanford large network dataset collection}.
\newblock \url{http://snap.stanford.edu/data/index.html}.

\bibitem{Barabasi-Science-99}
A.-L. B\'{a}rab\'{a}si and R.~Albert.
\newblock {Emergence of Scaling in Random Networks}.
\newblock {\em Science}, 286:509--512, 1999.

\bibitem{Borgs-WINE-10}
C.~Borgs, J.~Chayes, A.~T. Kalai, A.~Malekian, and M.~Tennenholtz.
\newblock {A Novel Approach to Propagating Distrust}.
\newblock In {\em WINE}, 2010.

\bibitem{Boshmaf-ACSAC-11}
Y.~Boshmaf, I.~Muslukhov, K.~Beznosov, and M.~Ripeanu.
\newblock {The Socialbot Network: When Bots Socialize for Fame and Money}.
\newblock In {\em ACSAC}, 2011.

\bibitem{Cao-NSDI-12}
Q.~Cao, M.~Sirivianos, X.~Yang, and T.~Pregueiro.
\newblock {Aiding the Detection of Fake Accounts in Large Scale Social Online
  Services}.
\newblock In {\em NSDI}, 2012.

\bibitem{Danezis-NDSS-09}
G.~Danezis and P.~Mittal.
\newblock {SybilInfer: Detecting Sybil Nodes using Social Networks}.
\newblock In {\em NDSS}, 2009.

\bibitem{Kerchove-SDM-08}
C.~de~Kerchove and P.~van Dooren.
\newblock {The PageTrust Algorithm: How to Rank Web Pages When Negative Links
  are Allowed?}
\newblock In {\em SDM}. SIAM, 2008.

\bibitem{Gao-IMC-10}
H.~Gao, J.~Hu, C.~Wilson, Z.~Li, Y.~Chen, and B.~Y. Zhao.
\newblock {Detecting and Characterizing Social Spam Campaigns}.
\newblock In {\em IMC}, 2010.

\bibitem{Guha-WWW-04}
R.~Guha, R.~Kumar, P.~Raghavan, and A.~Tomkins.
\newblock {Propagation of Trust and Distrust}.
\newblock In {\em WWW}, 2004.

\bibitem{Hanley-Radiology-1982}
J.~A. Hanley and B.~J. {McNeil}.
\newblock {The Meaning and Use of the Area under a Receiver Operating
  Characteristic {(ROC)} Curve}.
\newblock {\em Radiology}, 143, 1982.

\bibitem{Kunegis-WWW-09}
J.~Kunegis, A.~Lommatzsch, and C.~Bauckhage.
\newblock {The Slashdot Zoo: Mining a Social Network with Negative Edges}.
\newblock WWW, 2009.

\bibitem{Leskovec-KDD-06}
J.~Leskovec and C.~Faloutsos.
\newblock {Sampling from Large Graphs}.
\newblock In {\em SIGKDD}, 2006.

\bibitem{Panagiotopoulos-SybilFence-11}
S.~Panagiotopoulos, Q.~Cao, M.~Sirivianos, A.~Stavrou, C.~Liang, and X.~Yang.
\newblock {Quantifying the Cost of Sybil Attacks in Online Social Networks}.
\newblock \url{http://tinyurl.com/quantifying-sybil-cost}, 2011.

\bibitem{post-2011-bazaar}
A.~Post, V.~Shah, and A.~Mislove.
\newblock {Bazaar: Strengthening user reputations in online marketplaces}.
\newblock In {\em {NSDI}}, {2011}.

\bibitem{Tran-NSDI-09}
D.~N. Tran, B.~Min, J.~Li, and L.~Subramanian.
\newblock {Sybil-Resilient Online Content Rating}.
\newblock In {\em NSDI}, 2009.

\bibitem{Viswanath-SIGCOMM-10}
B.~Viswanath, A.~Post, K.~P. Gummadi, and A.~Mislove.
\newblock {An Analysis of Social Network-based Sybil Defenses}.
\newblock In {\em {ACM SIGCOMM}}, {2010}.

\bibitem{Yang-IMC-11}
Z.~Yang, C.~Wilson, X.~Wang, T.~Gao, B.~Y. Zhao, and Y.~Dai.
\newblock {Uncovering Social Network Sybils in the Wild}.
\newblock In {\em IMC}, 2011.

\bibitem{Yu-SP-08}
H.~Yu, P.~Gibbons, M.~Kaminsky, and F.~Xiao.
\newblock {SybilLimit: A Near-Optimal Social Network Defense Against Sybil
  Attacks}.
\newblock In {\em IEEE S\&P}, 2008.

\bibitem{SybilGuard-SIGCOMM-06}
H.~Yu, M.~Kaminsky, P.~B. Gibbons, and A.~Flaxman.
\newblock {SybilGuard: Defending Against Sybil Attacks via Social Networks}.
\newblock In {\em SIGCOMM}, 2006.

\bibitem{Ziegler-ISF-05}
C.-N. Ziegler and G.~Lausen.
\newblock {Propagation Models for Trust and Distrust in Social Networks}.
\newblock {\em Information Systems Frontiers}, 7, 2005.

\end{thebibliography}
